\documentclass[12pt]{article}
\usepackage{epsfig}
\newcommand{\nat}{Nature,~ }
\newcommand{\apjl}{Ap. J. Lett.,~}
\newcommand{\apj}{Ap. J., ~}
\newcommand{\physrep}{Physics Reports,~}

\def\beq{\begin{equation}}
\def\eeq#1{\label{#1}\end{equation}}
\def\eeqn{\end{equation}}

\def\beqa{\begin{eqnarray}}
\def\eeqa#1{\label{#1}\end{eqnarray}}
\def\eeqan{\end{eqnarray}}

\let\bar=\overbar

\def\Dslash{\not{\hbox{\kern-4pt $D$}}}
\def\dslash{\not{\hbox{\kern-2pt $\del$}}}

\def\msb{{\bar{\ssstyle M \kern -1pt S}}}

\def\Title#1{\begin{center} {\Large {\bf #1} } \end{center}}

\begin{document}

\Title{How to Construct a GRB Engine?}

\bigskip\bigskip


\begin{raggedright}

{\it Tsvi Piran and Ehud Nakar\index{Piran, T. Nakar, E.}\\
Racah Institute for Physics\\
The Hebrew University\\
Jerusalem, ISRAEL}
\bigskip\bigskip
\end{raggedright}

\section{Introduction}

Our understanding of Gamma-Ray Bursts  (GRBs) was revolutionized
during the last ten years. According to the generally accepted
Fireball model (see e.g. \cite{Pi99,Pi00}) the gamma-rays and
their subsequent multiwavelength afterglow are produced when an
ultra-relativistic flow is slowed down.   All current
observations, from prompt emission to late afterglow, from
$\gamma$-rays via X-ray optical and IR to radio, are consistent
with this model.

Observations of GRB host galaxies revealed the association of
(long) GRBs (there is no information on the positions of short
GRBs) with star forming Galaxies \cite{Fruchter99} and that these
(long) GRBs follow the star formation rate
\cite{Wijers98,Totani99,Blain00}. There is evidence (still
inconclusive) that some long bursts are associated with Supernovae
\cite{98bw,Bloom99,Reichart99,Galama00}. These observations
suggest that the progenitors of GRBs are massive stars. There are
some reasonable ideas how does a collapsing star
\cite{Woosley93,Pac98,MacFadyen_W99} or a merging binary
\cite{Eichler_LPS89} produce the required $\sim 10^{51}$ergs. But
it is not clear how does the GRB's ``inner engine" accelerates and
collimated the relativistic flow. Today this is the most
interesting (and most difficult) open question concerning GRBs.
This is also the most relevant question for this conference where
the possibility that transitions to strange stars power GRBs has
been  discussed \cite{Ouyed02}.

We summarize, here,  the known constrains on the ``inner engines"
of GRBs. These constrains arise mostly from the temporal structure
of GRBs. Several years ago Sari and Piran \cite{SP97} have shown
that variable GRBs can be generated only by internal
interactions\footnote{This interaction is usually considered as a
collisionless shock. However the exact nature of the interaction
is unimportant for most of our arguments.} within the flow. To
produce internal shocks the central engine must produce a long
and variable wind. This leads to a powerful {\bf NO GO} theorem:
\break {\bf Variable GRBs cannot be produced from a single
explosion.} \break This NO GO theorem rules out  explosive GRB
models that produce the relativistic flow in a single explosion.
Kobayashi et al \cite{KPS97} have shown that the observed
internal shocks light curve reflects almost directly the temporal
activity of the inner engine. This is the best direct evidence on
what is happening at the center of the GRB.

We review  the arguments leading to these conclusions. We also
discuss new observational results \cite{NK01,NK02a,NK02b} and a
new theoretical toy model \cite{NK02c} that explains these
observations within the internal shocks paradigm. This toy model
suggests that the ``inner engine" is producing a variable Lorentz
factor wind by modulating the mass ejection of a roughly constant
energy flow. The other alternative of modulating the energy of a
constant mass flow is ruled out. We conclude by summarizing the
various constrains on the ``inner engines". We leave to the
reader the task of examining the implications of these results to
his/hers favorite model\footnote{See \cite{NPK01} for a
discussion of the implications for possible accretion models}.

\section{Energetics and Beaming}

The most important factor in any model is the total amount of
energy that it releases. Redshift measurements have lead to
alarming estimates of more than $10^{54}$ergs in some bursts
\cite{Kulkarni99}. When factoring in the efficiency the
requirements exceed a solar rest mass energy.

However, these early estimates assumed isotropic emission. Jet
breaks in afterglow light curves lead to estimates of the beaming
factors. When those are taken into account we find a ``modest"
practically constant energy release of $\sim 10^{51}$ergs
\cite{Pi01a,Pi01b,PK01,Frailetal01}. This lower energy budget
allows for many possible models. At the same time it introduces an
additional requirement on the central engine. It has to collimate
the relativistic flow to narrow beaming angles (at times or order
of $1^o-2^o$).

We cannot  estimate directly the total energy released by the
``inner engine". However, here are two possible estimates:
$E_\gamma$, the energy released as $\gamma$-rays and $E_K$, the
kinetic energy during the adiabatic afterglow phase. Remarkably
both energies are comparable. This last observation implies that
the conversion efficiency of the initial relativistic kinetic
energy to $\gamma$-rays must be very high.

\section{Time Scales In GRBs - Observations}
\label{sec:time-obs}

\begin{figure}[htb]
\begin{center}
\epsfig{file=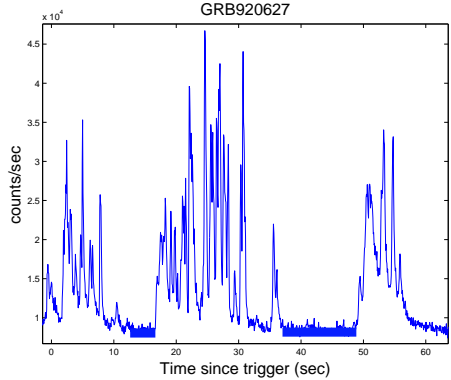,height=2in} \caption{The light curve of
GRB920627. The total duration of the burst is 52sec, while
typical pulses are 0.8sec wide. Two quiescent periods lasting
$\sim$10 seconds are marked by horizontal solid bold lines. }
\label{fig:variable}
\end{center}
\end{figure}

Most GRBs are highly variable.  Fig \ref{fig:variable}  depicts
the light curve of a typical variable GRB (GRB920627).  The
variability time scale, $\delta t$,  is determined by the width
of the peaks. $\delta t$  is much shorter (in some cases  by a
more than a factor of 100) then $T$, the duration of the burst.
Variability on a time scale of milliseconds is seen in some long
bursts   \cite{NK02b}. However, not all bursts are variable. We
stress that  our discussion applies to variable bursts and it is
not applicable to the small subset of smooth ones.

A  comparison of the pulse width distribution and the pulse
separation, $\Delta t$ , distribution, reveals an excess of long
intervals \cite{NK01,NK02a}. These long intervals can be
classified as quiescent periods \cite{Ramirez-Ruiz-Melroni},
relatively long periods of several dozen seconds with no
activity. When excluding the  quiescent periods we
\cite{NK01,NK02a} find that both distributions are lognormal with
a comparable parameters: The average pulse interval, $\bar \Delta
t = 1.3sec$  is larger by a factor 1.3 then the average pulse
width $\bar \delta t= 1sec$. One also finds that the pulse widths
are correlated with the preceding interval \cite{NK01,NK02a}.

\begin{figure}[htb]
\begin{center}
\epsfig{file=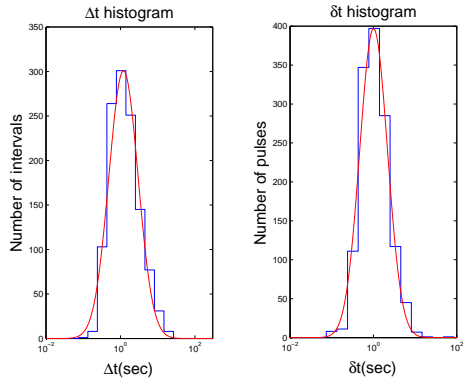,height=2in} \caption{The pulse
width distribution (right) and the distribution of intervals
between pulses (left) (from \cite{NK02a}).}
\label{fig:distributions}
\end{center}
\end{figure}

The results described so far are for long bursts.  the
variability of short ($T<2$sec) bursts is more difficult to
analyze. The duration of these bursts is closer to the limiting
resolution of the detectors. Still we  find that most ($\sim
66\%$) short bursts are variable with $\delta t/T < 0.1$
\cite{NK02b}.

\section{Time Scales In GRBs - Theory}

\begin{figure}[htb]
\begin{center}
\epsfig{file=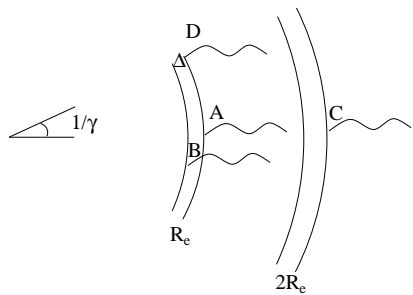,height=1.5in} \caption{Different time
scale from a relativistic expanding shell in terms of the arrival
times ($t_i$) of various photons:  $t_{ang} = t_D-t_A$, $t_{rad} =
t_C-t_A$ and $T_\Delta = t_B-t_A$.} \label{fig:times}
\end{center}
\end{figure}

Consider a spherical relativistic emitting shell with a radius
$R$, a width $\Delta$  and a Lorentz factor $\Gamma$.  This can
be a whole spherical shell or a spherical like section of a jet
whose opening angle $\theta$ is larger than $\Gamma^{-1}$.
Because of relativistic beaming an observer would observe
radiation only from a region of angular size $\Gamma^{-1}$.
Photons emitted by matter moving directly towards the observer
(point A in Fig. \ref{fig:times}) will arrive first. Photons
emitted by matter moving at an angle $\Gamma^{-1}$ (point D in
Fig. \ref{fig:times}) would arrive after $t_{ang} =
R/2c\Gamma^2$. This is also the time of arrival of photons
emitted by matter moving directly towards the observer but
emitted at $2R$ (point C in Fig. \ref{fig:times}). Thus, $t_{rad}
\approx t_{ang}$ \cite{SP97,Fenimore96}. This coincidence is the
first part of the NO GO theorem that rules out single explosions
a sources of GRBs.

At a given point particles are continuously accelerated and emit
radiation as long as the shell with a width $\Delta$ is crossing
this point. The photons emitted at the front of this shell will
reach the observer a time  $T_\Delta = \Delta /c$ before those
emitted from the rear (point B in Fig. \ref{fig:times}). In fact
photons are emitted slightly longer as it takes some time for the
accelerated photons to cool.  For most reasonable parameters the
cooling time is much shorter from the other time scales
\cite{SNP96} and we ignore it hereafter.

Light curves are divided to two classes according to the ratio
between $T_\Delta $ and $t_{ang} \approx t_{rad}$. The emission
from different angular points smoothes the signal on a time scale
$t_{ang}$.   If $T_\Delta \le t_{ang}\approx t_{rad}$ the
resulting burst will be smooth with a width $t_{ang}\approx
t_{rad}$. The second part of the NO GO theorem follows from the
hydrodynamics of external shocks. Sari and Piran \cite{SP97} have
shown that for external shocks $\Delta/c \le R/c \Gamma^2
\approx  t_{rad} \approx t_{ang}$. External shocks can produce
only smooth bursts!

\begin{figure}[htb]
\begin{center}
\epsfig{file=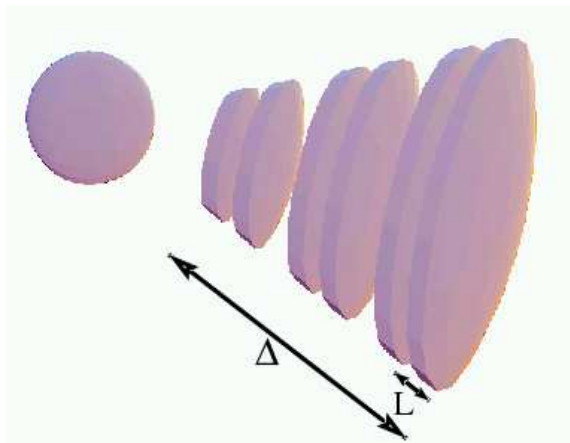,height=2.3in,width=3in}
\caption{The internal shocks model (from \cite{Sari_PhD}). Faster
shells collide with slower ones and produce the observed $\gamma$
rays. The variability time scale is $L/c$ while the total
duration of the burst is $\Delta/c$ .} \label{fig:internal_shocks}
\end{center}
\end{figure}

A necessary condition for the production of a variable light
curve is that $T_\Delta =  \Delta/c >   t_{ang}$. This can be
easily satisfied within internal shocks (see Fig
\ref{fig:internal_shocks}). Consider an ``inner engine" emitting
a relativistic wind active over a time $T_\Delta =\Delta/c$
($\Delta$ is the overall width of the flow in the observer
frame). The source is variable on a scale $L /c$. The internal
shocks will take place at $R_s \approx L \Gamma^2$. At this place
the angular time and the radial time satisfy: $t_{ang} \approx
t_{rad} \approx L/c $. Internal shocks continue as long as the
source is active, thus the overall observed duration $T =
T_\Delta$ reflects the time that the ``inner engine" is active.
Note that now $t_{ang} \approx L/c < T_\Delta$ is trivially
satisfied. The observed variability time scale in the light
curve, $\delta t$,  reflects the variability of the source $L/c$.
While the overall duration of the burst reflects the overall
duration of the activity of the ``inner engine".

Numerical simulations  \cite{KPS97} have shown that not only the
time scales are preserved but the source's temporal behaviour is
reproduced on an almost  one to one basis in the observed light
curve. We will return to this point in section \ref{New} in which
we describe a simple toy model that explains this result.

\section{Caveats and Complications}

Clearly the way to get around the NO GO theorem is if $t_{ang} <
t_{rad}$. In this case one can identify $t_{rad}$ with the
duration of the burst and $t_{ang}$ as the variability time
scale. The observed variability would require in this case that:
$t_{ang}/t_{rad} = \delta t /T$.

\begin{figure}[htb]
\begin{center}
\epsfig{file=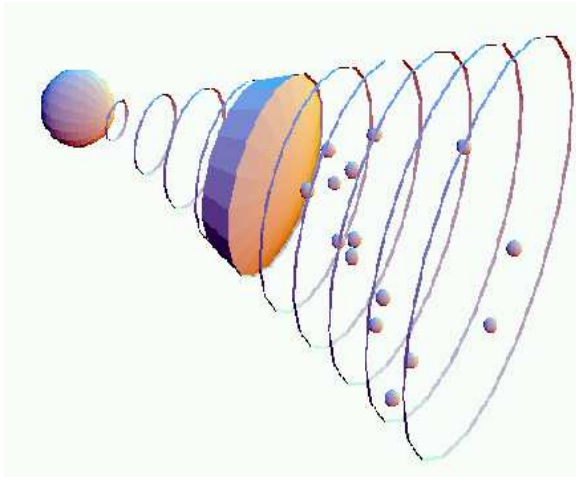,height=2.5in,width=3in} \caption{The
clumpy ISM model (from \cite{Sari_PhD}). Note the small covering
factor and the resulting ``geometrical" inefficiency.}
\label{fig:clumps}
\end{center}
\end{figure}

One can imagine an  inhomogeneous external medium which is clumpy
on a scale $d \ll R/\Gamma$ (see Fig \ref{fig:clumps}). Consider
such a clump located at an angle $\theta \sim \Gamma^{-1}$ to the
direction of motion of the matter towards the observer. The
resulting angular time, which is the difference in arrival time
between the first and the last photons emitted from this region
would be:$\sim d/c \Gamma $. Now $t_{ang} < t_{rad}$ and it seems
that one can get around the NO GO theorem.

Sari and Piran \cite{SP97} have shown that such a configuration
would be extremely inefficient. This  third component of the NO GO
theorem rules out this caveat. The observations limit the size of
the clumps to $d \approx c \Gamma \delta t$ and the location of
the shock to  $R \approx c T \Gamma^2 $. The number of clumps
within the observed angular cone with an opening angle
$\Gamma^{-1}$ equals the number of pulses which is of the order
$T/\delta t$. The covering factor of the clumps can be directly
estimated in terms of the observed parameters by multiplying the
number of clumps $T/\delta t$ times their area $d^2= (\delta t
\Gamma)^2$ and dividing by the cross section of the cone
$(R/\Gamma)^2$. The resulting covering factor equals $ \delta t
/T \ll 1$.  The efficiency of conversion of kinetic energy to
$\gamma$-rays in this scenario is smaller than this covering
factor which for a typical variable burst could be smaller than
$10^{-2}$.

We turn now to several attempts to find a way around this result.
We will not discuss here the feasibility of the suggested models
(namely is it likely that the surrounding matter will be clumpy
on the needed length scale \cite{Dermer_Mitman99}, or can an
inner engine eject ``bullets" \cite{Begelman99} or ``cannon
balls" \cite{Dar_DeRujula99} with an angular width of $\sim
10^{-2}$ degrees and what keeps these bullets so small even when
they are shocked and heated). We examine only the question
whether the observed temporal structure can arise within these
models.

\subsection{External Shocks on a Clumpy Medium}

Dermer and Mitman \cite{Dermer_Mitman99} claim that the simple
efficiency argument of Sari and Piran \cite{SP97} was flawed.
They point out that if the direction of motion of a specific blob
is almost exactly towards the observer the angular time would be
of order $d^2/cR$. This is narrower by a factor $d\Gamma/R$ than
the angular time across the same blob that is located at a
typical angle of $\Gamma^{-1}$. These special blobs would produce
strong narrow peaks and will form a small region along a narrow
cone with a larger covering factor. Dermer and Mitman
\cite{Dermer_Mitman99} present a numerical simulation of light
curves produced by external shocks on a clumpy inhomogenous
medium with $\delta t/ T \sim 10^{-2} $ and  efficiencies of up to
$\sim 10$\%.

A detailed analysis of the light curve poses, however, several
problems for this model.  While this result is marginal for bursts
with $\delta t/T \sim 10^{-2}$ it is insufficient for bursts with
$\delta t /T \sim 10^{-3}$.   Variability on a time scale of
milliseconds is observed \cite{NK02b} in many long  GRBs (namely
$\delta t / T $ can be as samll as $10^{-4}$.). Moreover, in this
case we would expect that earlier pulses (that arise from blobs
along the direction of motion) would be narrower than latter
pulses. This is in a direct contradiction with the observations
\cite{Fenimore99}. Finally there is no reason to expect the
observed similarity between the pulse width and the pulse
separation and the correlation between the pulse width and the
preceding interval in this model.

\subsection{The Shot-Gun and the Cannon Ball Models}

\begin{figure}[htb]
\begin{center}
\epsfig{file=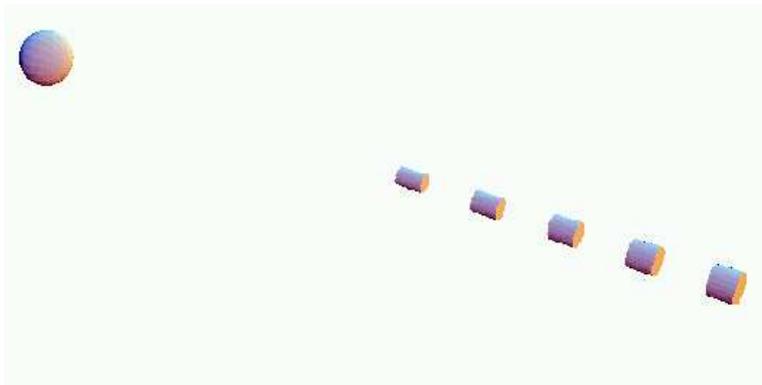,height=2in,width=4in} \caption{The
shot-gun or the cannon-ball  models (from \cite{Sari_PhD}). The
inner engine emits narrow ``bullets" that collide with the ISM.}
\label{fig:bullets}
\end{center}
\end{figure}

Heinz and Begelman  \cite{Begelman99} suggested  that the ``inner
engine" operates as a shot-gun emiting multiple narrow bullets
with an angular size much smaller than $\Gamma^{-1}$ (see Fig
\ref{fig:bullets}). These bullets do not spread while propagating
and they  are slowed down rapidly by an external shock with a
very dense circumburst matter. The pulses width is $t_{ang}$ or
the slowing down time while the duration of the burst is
determined by the time that the ``inner engine" emits the
bullets. While the cannon ball model of Dar and De Rujulla
\cite{Dar_DeRujula99} is drastically different from a physical
point of view it is rather similar in terms of its temporal
features. Hence the following remarks apply to this model as well.

This model satisfies our NO GO theorem in the sense that also here
the burst is not produced by a single explosion. Moreover the
observed light curve represents also here the temporal activity
of the source. However, it is based on external shocks.

The most serious problem  is the fact that the width of the
pulses here is determined by the angular time or the hydrodynamic
time (which in turn depends on the external density profile of the
circumburst matter) while the intervals between the pulses depend
on the activity of the inner engine. There is no reason why the
two distributions will be similar and why there should be a
correlation between them.

\section{An Internal Shocks Toy Model}
\label{New}
 The discovery  \cite{NK01,NK02a} that the distribution
of pulse widths and pulse separations are comparable  and that
there is a correlation between the pulse width and the preceding
interval provides an independent evidence in favor of the
internal shocks model. Furthermore it suggests that the different
shells emitted by the internal engine are most likely  ``equal
energy" rather than ``equal mass" shells.

The similarity between the pulse width and the pulse separation
distribution  and the correlation is extremely unlikely  to arise
from a single shell passing thought a  random distribution of
clumps that surround the ``central engine". In this case the
arrival time of individual pulses will depend on the position of
the emitting clumps relative to the observers. Two following
pulses could arise from two different clumps that are rather
distant from each other. There is no reason why the pulses and
intervals should be correlated in any way. A similar problem
arises in the ``shot-gun" or the ``cannon-ball" models in which
the pulse duration is determined by one parameter and the
separation by another.

Both  features arise naturally within the internal shocks model
\cite{NK02c} in which both the pulse duration  and the separation
between the pulses are determined by the same parameter. We
outline here the main arguments showing that. Consider two shells
with a separation $ L $. The slower outer shell Lorentz factor is
$ \Gamma _{1}=\Gamma $ and the inner faster shell Lorentz factor
is $ \Gamma _{2}=a\Gamma  $ ($ a>2 $ for an efficient collision),
both in the observer frame. The shells' are ejected at $ t_{1} $
and $ t_{2}\approx t_{1}+L/c$. The collision takes place at a
radius $ R_s\approx 2\Gamma ^{2}L $ (Note that $ R_s $ does not
depend on $ \Gamma _{2} $). The emitted photons from the
collision will reach the observer at time (omitting the photons
flight time, and assuming transparent shells):

\begin{equation}
\label{to} t_{o} \approx t_{1}+R_s/( 2c\Gamma ^{2})\approx
t_{1}+L\approx t_{2} \ .
\end{equation}
The photons from this pulse are observed almost simultaneously
with a (hypothetical) photon that was emitted from the ``inner
engine'' together with the second shell (at $ t_{2} $). This
explains why Kobayashi et al \cite{KPS97} find numerically that
for internal shocks the observed light curve replicates the
temporal activity of the source.

Consider now  four shells emitted at times $ t_{i} $ ($ i=1,2,3,4
$) with a separation of the order of $ L $ between them. Assume
that there are two collisions - between the first and the second
shells and between the third and the fourth shells. The first
collision will be observed at $ t_{2} $ while the second one will
be observed at $ t_{4} $. Therefore, $ \Delta t\approx
t_{4}-t_{2}\approx 2L/c $. Now assume a different collision
scenario, the second and the first shells collide, and afterward
the third shell takes over and collide with them (the forth shell
does not play any roll in this case). The first collision will be
observed at $ t_{2} $ while the second one will  be observed at $
t_{3} $. Therefore, $ \Delta t\approx t_{3}-t_{2}\approx L/c. $
Numerical simulations \cite{NK02c} show that more then 80\% of
the efficient collisions follows one of the two scenarios
described above. Therefore we can conclude:
\begin{equation}
\Delta t\approx L/c \ . \label{separation}
\end{equation}
Note that this result is independent of the shells' masses.

The pulse width is determined by the angular time (ignoring the
cooling time): $ \delta t=R_s/(2c\Gamma ^{2}_{s}) $ where $
\Gamma _{s} $ is the Lorentz factor of the shocked emitting
region. If the shells have an equal mass ($ m_{1}=m_{2} $) then $
\Gamma _{s}=\sqrt{a}\Gamma $ while if they have equal energy ($
m_{1}=am_{2} $) then $ \Gamma _{s}\approx \Gamma  $. Therefore:
\begin{equation}
\delta t \approx
  \left\{ \begin{array}{r@{\quad\quad}l}
    R_s/2a\Gamma^{2}c\approx L/ac & \rm{equal \  mass}, \\
    R_s/2\Gamma ^{2}c \approx L/c  & \rm {equal \ energy}.
    \end{array} \right .
    \label{width}
\end{equation}
The ratio of the  Lorentz factors $ a $, determines  the
collision's efficiency. For efficient collision the variations in
the shells Lorentz factor (and therefore  $ a $) must be large.

It follows from Eqs. \ref{separation} and \ref{width} that for
equal energy shells the $ \Delta t $-$ \delta t $ similarity and
correlation arises naturally from the reflection of the shells
initial separation in both variables. However, for  equal mass
shells $ \delta t $ is shorter by a factor of $a$ than $ \Delta t
$. Since  $ a $ has a large variance this would wipes off the $
\Delta t $-$ \delta t $ correlation. This suggests that equal
energy shells are more likely to  produce the observed light
curves.

\section{Conclusions}

We cannot provide a recipe for a GRB ``inner engine". However we
can list the specifications of this engine (for a long variable
GRB). It must satisfy the following conditions:
\begin{itemize}
\item It should accelerate $\sim 10^{51}$ergs to a variable relativistic
flow with $\Gamma > 100$.
\item It should collimate this flow, with a varying degree of
collimation (up to $1^o$).
\item It should be active from several seconds  up to several  hundred
seconds (according to the duration of the observed burst).
\item It should vary on a time scale of seconds or less (corresponding to the
duration of a typical pulse within the burst).
\item Different shells of matter should have a comparable energy
and their different Lorentz facors should arise due to a
modulation of the accelerated mass.
\item At times the engine should stop for several dozen
seconds (resulting in a quiescent periods).
\end{itemize}

Before concluding we stress that these specification are for long
variable burst (which compose the majority of the observed
bursts). Many of these (but not all) apply also to short variable
bursts (about two thirds of the short bursts). These
specifications do not apply to smooth bursts (either short or
long ones).

\bigskip
This research was supported by a grant from the US-Israel
Binational Science  Foundation.

\def\Discussion{
\setlength{\parskip}{0.3cm}\setlength{\parindent}{0.0cm}
     \bigskip\bigskip      {\Large {\bf Discussion}} \bigskip}
\def\speaker#1{{\bf #1:}\ }
\def\endDiscussion{}

\end{document}